\documentclass[journal]{IEEEtran}
\IEEEoverridecommandlockouts
\usepackage{mathrsfs}
\usepackage{amsfonts}
\usepackage{ifpdf}
\usepackage{cite}

\usepackage{xcolor}

\ifCLASSINFOpdf
\usepackage[pdftex]{graphicx}
\else \fi
\usepackage[cmex10]{amsmath}
\usepackage{array}
\usepackage{mdwmath}
\usepackage{mdwtab}
\usepackage{eqparbox}
\usepackage[tight,footnotesize]{subfigure}
\usepackage{algorithmic}
\usepackage{algorithm}
\usepackage{amsmath}
\usepackage{epsfig}
\usepackage{ae}
\usepackage{amssymb}
\usepackage{times}
\usepackage{amsthm}
\usepackage{amsmath}   

\begin{document}

\title{Green Heterogeneous Cloud Radio Access Networks: Potential Techniques, Performance Tradeoffs, and Challenges}

\author{
         Yuzhou~Li,
         Tao~Jiang,
         Kai~Luo,
         and~Shiwen~Mao
\thanks{Y. Li, T. Jiang, and K. Luo are with Huazhong University of Science and Technology and S. Mao is with Auburn University.}
}
\maketitle
\IEEEpeerreviewmaketitle
\begin{abstract}
As a flexible and scalable architecture, heterogeneous cloud radio access networks (H-CRANs) inject strong vigor into the green evolution of current wireless networks.  But the brutal truth is that energy efficiency (EE) improves at the cost of other indexes such as spectral efficiency (SE), fairness, and delay. It is thus important to investigate performance tradeoffs for striking flexible balances between energy-efficient transmission and excellent quality-of-service (QoS) guarantees under this new architecture. In this article, we first propose some potential techniques to energy-efficiently operate H-CRANs by exploiting their features. We then elaborate the initial ideas of modeling three fundamental tradeoffs, namely EE-SE, EE-fairness, and EE-delay tradeoffs,  when applying these green techniques, and present open issues and challenges for future investigations. These related results are expected to shed light on green operation of H-CRANs from adaptive resource allocation, intelligent network control, and scalable network planning.
\end{abstract}
\section{Introduction}
\subsection{Background and Motivation}

The dramatic increase in the number of smart phones and tablets with ubiquitous broadband connectivity has triggered an explosive growth in mobile data traffic \cite{CiscoMobileDataTrafficForecast_2016_2021}. Cisco forecasts that, the amount of global mobile data traffic will increase 7-fold from 2016 to 2021 and its majority is generated by energy-hungry applications such as mobile video \cite{CiscoMobileDataTrafficForecast_2016_2021}. This is also referred to as the well-known $1000\times$ data challenge in cellular networks. Meanwhile, the number of devices connected to the global mobile communication networks will reach 100 billion in the future and that of mobile terminals will surpass 10 billion by 2020 \cite{5G_VisionAndRequirements}.

Although unprecedented opportunities for the development of wireless networks brought by the massive traffic amount and connected devices, a concomitant crux is that this growth skyrockets the energy consumption (EC) and greenhouse gas emissions in the meantime. From statistical data, the information and communication technology (ICT) industry is responsible for $2\%$ of world-wide $\text{CO}_2$ emissions and $2\%$-$10\%$ of global EC, of which more than $60\%$ is directly attributed to radio access networks (RANs) \cite{Footprint_of_EcologicalAndEconomic_CM2011}. For this regard, 5G wireless communication networks are anticipated to provide spectral and energy efficiency growth by a factor of at least 10 and 10 times longer battery life of connected devices \cite{5G_VisionAndRequirements}.

\subsection{Concept of H-CRANs}
To meet the $1000\times$ data challenge, heterogeneous networks (HetNets), composed of a diverse set of small cells (e.g., microcells, picocells, and femtocells) overlaying the conventional macrocells, have been introduced as one of the most promising solutions \cite{5G_VisionAndRequirements}. However, the ubiquitous deployment of HetNets is accompanied by the following shackles:
\begin{itemize}
  \item \textbf{Severe interference}. The spectrum re-use among cells incurs severe mutual interference, which may significantly reduce the expected system spectral efficiency (SE) and also decrease the network energy efficiency (EE).
  \item \textbf{Unsatisfactory EE}. The densely-deployed small cells lead to an escalated EC and thus a reduced EE, and also increases capital expenditures (CAPEX) and operational expenditures (OPEX).
  \item \textbf{No computing-enhanced coordination centers}. There are no centralized units with strong computing abilities to globally coordinate multi-tier interference and execute cross-RAN optimization, which dramatically limits cooperative gains among cells.
  \item \textbf{Inflexibility and unscalability}. Fragmented base stations (BSs) result in inflexible and unscalable network control and operations, thus leading to redundant network planning and inconvenient network upgrade.
\end{itemize}

To overcome these challenges faced by HetNets, cloud RANs (C-RANs),  new centralized cellular architectures armed with powerful cloud computing and virtualization techniques, have been parallelly put forward to coordinate interference and manage resources across cells and RANs \cite{CRAN_WhitePaper}. In C-RANs, a large number of low-cost low-pwer remote radio heads (RRHs), connecting to the baseband unit (BBU) pool through the fronthaul links, are randomly deployed to enhance the wireless capacity in the hot spots. Consequently, the combination of HetNets and C-RANs, known as heterogeneous C-RANs (H-CRANs), becomes a potential solution to support both spectral- and energy-efficient transmission.

\subsection{Green H-CRANs}
As mentioned above, one of the main missions for H-CRANs from their birth is to construct eco-friendly and cost-efficient wireless communication systems. Benefiting from H-CRANs' global coordination ability, many promising techniques, such as joint processing/allocation, traffic load offloading, energy balance, self-organization, and adaptive network deployment, can be applied in these scenarios for energy-efficient transmissions. Unfortunately, the network EE improves usually at the cost of the performance of other technique metrics, such as SE, fairness, and delay, all of which however are equally important as EE to guarantee users' quality-of-service (QoS). That is, there are EE-SE, EE-fairness, and EE-delay tradeoffs. It is thus interesting to investigate these performance tradeoffs in H-CRANs for establishing rules to flexibly balance the network EE and users' QoS demands when greening H-CRANs.

Compared with existing works (e.g. \cite{RecentAdvances_CRAN_Survey2016}) on the system architecture or radio resource management (RRM) mainly in terms of EE and SE, this article focuses on the green evolution of H-CRANs, and particularly investigates it from the perspective of EE-SE, EE-fairness, and EE-delay tradeoffs instead of the indexes themselves. To reach our targets, we organize the remainder of this article as follows. In Section~\ref{Section:Architecture_HCRANs}, we first simply review the architecture of H-CRANs and then exploit their features to propose three potential techniques for green H-CRANs. Section~\ref{Section:TradeoffTheories} introduces the possible methods to depict these tradeoffs and also provides corresponding challenges and open problems when applying these proposed techniques. We conclude the article in Section~\ref{Section:Conlusions}.

\section{Architecture of H-CRANs and Potential Green Techniques} \label{Section:Architecture_HCRANs}
In C-RANs, the idea of dividing conventional cellular BSs into two parts of BBUs and RRHs is introduced. BBUs are then integrated into centralized BBU pools, where cloud computing and virtualization techniques are implemented to enhance computational ability and to virtualize network function. BBUs are responsible for resource control and signal processing, while RRHs for information radiation and reception, with their interconnection via dedicated transport networks. Thus, the cloud-computing-enhanced centralized BBU pools facilitate cross-cell and cross-RAN information sharing, which paves the path for global resource optimization adapting to network conditions (e.g., channel conditions, interference strength, traffic loads, and so on). H-CRANs absorb this architecture in C-RANs and maintain macro BSs (MBSs) and small-cell BSs (SBSs) in HetNets to support both global control and seamless communications.

\subsection{Architecture of H-CRANs}
\begin{figure}[t]
\centering \leavevmode \epsfxsize=3.2in  \epsfbox{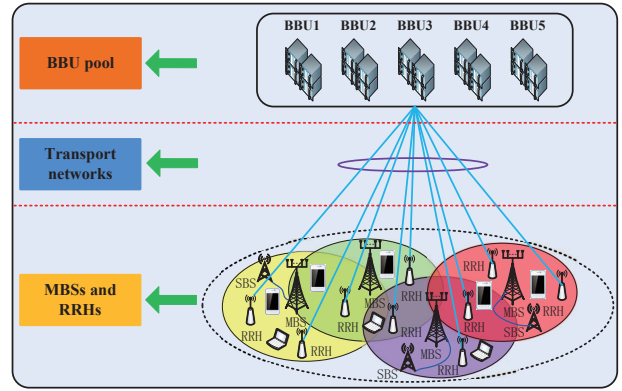}
\centering \caption{The architecture of H-CRANs.} \label{Fig:H_CRAN}
\end{figure}

As shown in Fig.~\ref{Fig:H_CRAN}, H-CRANs are composed of three functional modules.
\begin{enumerate}
  \item \textbf{Real-time virtualization and cloud-enhanced BBU pool}. Equipped with powerful virtualization techniques and strong real-time cloud computing ability, BBU pools integrate independent BBUs scattered in cells.
  \item \textbf{High-reliability transport networks}. RRHs are connected to BBUs in the BBU pool via high-bandwidth low-latency fronthaul links such as optical transport networks. The data and control interfaces between the BBU pool and MBSs are S1 and X2, respectively \cite{HCRAN_EE_Perspective_IWC2014}.
  \item \textbf{MBSs, SBSs, and RRHs}. In H-CRANs, multiple access points (APs), e.g., MBSs, SBSs, and RRHs, coexist. MBSs are deployed mainly for network control and mobility performance improvement, e.g., decreasing handover times to avoid Ping-Pong effects for high-mobility users. SBSs and RRHs are geographically distributed within cells close to users to increase capacity and decrease transmit power in the meantime.
\end{enumerate}

In H-CRANs, the function separation between BBUs and RRHs, the decoupling between control and data planes, and the cloud-computing-enhanced centralized integration of BBUs facilitate efficient management of densely-deployed mobile networks. For example, the operators only need to install new RRHs and connect them to the BBU pool to expand network coverage and improve network capacity. Moreover, flexible software solutions can be easily implemented under this architecture. For instance, the operators can upgrade RANs and support multi-standard operations only through software update by deploying software defined radio (SDR).

\subsection{Potential Techniques for Green H-CRANs}
The four revolutionary changes, i.e., function separation, control-data decoupling, centralized architecture, and cloud-computing-enhanced processing, make H-CRANs significantly different from existing 2G, 3G, and 4G wireless networks. By exploiting these features, it is possible to construct H-CRANs flexible in network management, adaptive in network control, and scalable in network planning. As a result, energy-efficient operation of H-CRANs without a significant loss in other indexes such as SE, fairness, and delay can be achieved.

\textbf{1) Joint Resource Optimization across RRHs and RANs}

In H-CRANs, each BBU first collects its individual network conditions and then shares this information within the BBU pool. As a result, this distributed-collection centralized-control architecture, further enhanced by virtualization techniques and cloud computing, enables efficient transmission/reception cooperation across RRHs and convenient global control across RANs. Consequently, the existing cooperative techniques, such as coordinated multi-point (CoMP) transmission, enhanced inter-cell interference coordination (eICIC), and interference alignment (IA), can be readily implemented in H-CRANs. All these techniques are self-contained in theory but have rarely been applied to conventional cellular networks because of difficulties in sharing and handling global network information.

\begin{figure}[t]
\centering \leavevmode \epsfxsize=3.5in  \epsfbox{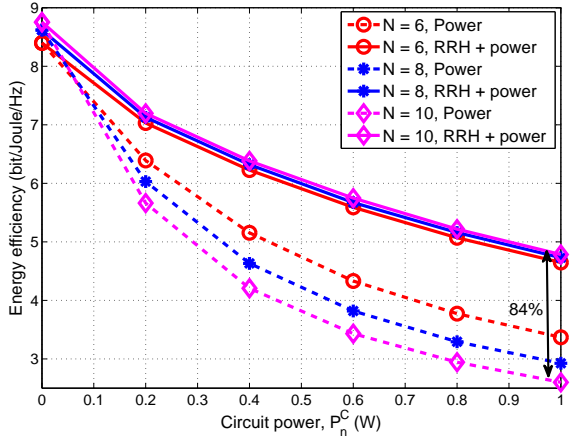}
\centering \caption{An example: EE variation with the circuit power of each RRH, denoted by $P_n^c$, in downlink H-CRANs, where a MBS, $N$ RRHs, and 16 users are included. In this example, we maximize the network EE by optimizing RRH operation and power allocation subject to constraints of users' minimum rate requirements of $R^{\text{req}} = 2$ bits/Hz.} \label{Fig:EE_VS_CircuitPower}
\end{figure}

As introduced above, multi-RANs and multi-APs with different coverage and functions are deployed in H-CRANs. As a result, unlike traditional single-mode terminals communicating only through a RAN's AP, multi-mode terminals could send and receive data concurrently through multiple of them. This indicates H-CRANs with a new characteristic of network diversity, which can be exploited to design user association strategies. By this, traffic load distributions among RANs and APs can be well balanced, which in turn affects the working states of RANs and resource optimization, and thus affects network interference and EE.

Moreover, under this new centralized architecture, the network EE can be further improved by incorporating more resource allocation dimensions (e.g., power allocation, subcarrier assignment, user association, and RRH operation) into the formulations. Fig.~\ref{Fig:EE_VS_CircuitPower} shows that joint optimization of RRH operation and power allocation improves EE by up to 84\% compared with the power-allocation-only algorithm in downlink H-CRANs. Thus, through the aforementioned joint resource optimization and network-diversity-aware user association, significant improvement in EE and reduction in EC can be achieved.

\textbf{2) Large-scale MBS and SBS Deployment}

Compared to the transmit power, the overall static power consumption by MBSs and SBSs, composed of cooling and circuit power, are usually much larger \cite{GreenCommunications_ConceptReality_IWC2012}. For example, a typical UMTS base station consumes 800--1500W with RF output power of 20--40W. As a result, under the constraints of basic coverage requirements, the deployment of MBSs and SBSs, characterized by the distance between two MBS sites and the number of SBSs per site, affects the area power consumption (APC) and the area SE (ASE) significantly in H-CRANs. The general purpose of large-scale MBS and SBS deployment is to macroscopically plan an appropriate number of BSs to support users' demands for energy saving by avoiding the static power consumption.

\begin{figure}[t]
\centering \leavevmode \epsfxsize=3.5in  \epsfbox{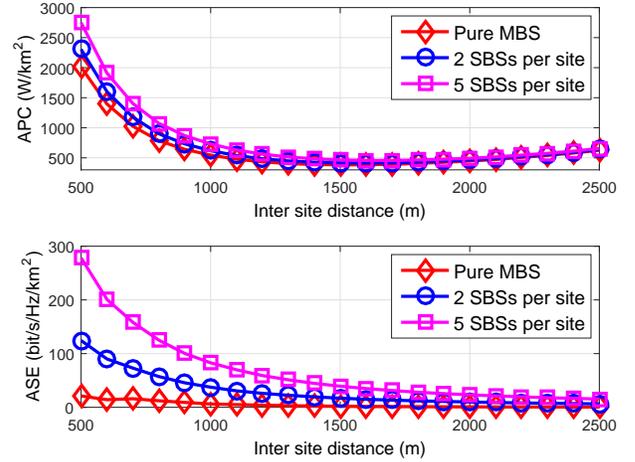}
\centering \caption{An example: The APC and ASE versus the inter site distance subject to a 95\% coverage constraint. In the figure, we adopt the practical models for the BS power consumption given by $P^{\text{tot}} = ap^{\text{tx}} + b$, where $a_{\text{MBS}} = 22.6$, $b_{\text{MBS}} = 412.4$ W, $a_{\text{SBS}} = 5.5$, and $b_{\text{SBS}} = 32$ W (note that SBSs refer to micro BSs in the figure) \cite{PowerConsumptionModelandParameters}.} \label{Fig:APC_ASE_Versus_ISD}
\end{figure}

Intuitively, the APC will sharply decrease if we reduce the number of MBSs, i.e., increase the inter site distance. Meanwhile, the ASE will also decrease, because the increased inter site distance reduces the spectrum re-use. Similarly, the number of SBSs deployed in each site will also affect the APC and the ASE. As an example, Fig.~\ref{Fig:APC_ASE_Versus_ISD} clearly shows the significant impacts of the configuration of MBSs and SBSs on the APC and ASE under practical parameter settings. Therefore, we need careful network planning from a large-scale perspective to flexibly balance these two metrics and to conveniently upgrade the system.

\textbf{3) Load-Aware RRH Operations}

The so-called worst-case network planning philosophy has been widely adopted to guarantee users' QoS even during peak traffic periods in conventional cellular networks. However, mobile traffic loads usually vary in both spatial and temporal domains, which is referred to as the tidal phenomenon. Specifically, the fraction of time when the traffic is below 10\% of the peak during a day is about 30\%  on weekdays and 45\% on weekends \cite{DynamicTrafficLoads_CellularNetwork}. As a result, a large number of RRHs are extremely under-utilized in the cases of dense deployment in H-CRANs during off-peak periods. But RRHs still consume circuit power even with little or no activity. Consequently, a significant waste of EC and a sharp decrease in EE will be resulted if RRHs are underutilized but still activated. Thus, apart from the aforementioned spatial deployment, energy conservation can also be achieved by exploiting temporal traffic variations. For the fixed deployment, we can adopt load-aware network control in H-CRANs to perform on/off operations of RRHs adapting to spatial and temporal traffic amounts to improve EE.

\begin{figure}[t]
\centering \leavevmode \epsfxsize=3.5in  \epsfbox{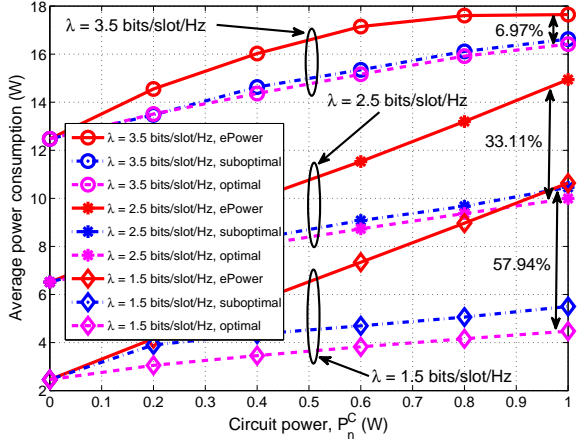}
\centering \caption{An example: Average power consumption with the circuit power of each RRH, denoted by $P_n^c$, under different traffic arrival rates $\boldsymbol \lambda$ in downlink H-CRANs, where a MBS, 8 RRHs, and 12 users are included. In this example, we jointly optimize RRH operation and power allocation to maximize the network EE considering stochastic and time-varying traffic arrivals.} \label{Fig:PowerConsumption_VS_CircuitPower}
\end{figure}

As an example, we consider a downlink H-CRAN to show the impacts of load-aware RRH on/off operations on energy expenditure. Specifically, we jointly optimize RRH operation and power allocation to maximize the network EE with stochastic and time-varying traffic arrivals taken into account. Two algorithms, denoted by the optimal and suboptimal, are developed to solve the problem. Fig.~\ref{Fig:PowerConsumption_VS_CircuitPower} shows that the proposed algorithms can dramatically reduce the energy consumption compared to the algorithm without RRH operation (i.e., only optimizing power allocation), denoted by ePower, especially in light and middle traffic states (up to a 58\% gain in light traffic states when the traffic arrival rate $\boldsymbol \lambda = 1.5$ bits/slot/Hz).

\section{Performance Tradeoffs and Challenges for Green H-CRANs} \label{Section:TradeoffTheories}
Leveraging the proposed potential green techniques in H-CRANs, it is then of importance to explore the key theories that support ubiquitous energy-efficient transmission and meanwhile provision satisfactory QoS for users. Among them, performance tradeoffs deserve significant consideration \cite{Funda_Tradeoff_YeLi}.

Apart from the widely studied deployment efficiency-EE, EE-SE, bandwidth-power, and delay-power tradeoffs \cite{Funda_Tradeoff_YeLi}, there are two additional fundamental tradeoffs, EE-fairness and EE-delay tradeoffs. This section elaborates the ideas of modeling these two tradeoffs, analyzes challenges and open problems, and provides some possible solutions. Since H-CRANs originally are designed to enhance the network SE and thus the wireless capacity as well, we thus also review the key concepts and present challenges associated with the EE-SE tradeoff under this new architecture.
\subsection{EE-SE Tradeoff}
Vast existing research falls into this direction due to the following reasons. The traditional indexes EC and SE measure how small the amount of energy is needed to satisfy users' QoS and how efficiently a limited spectrum is utilized, respectively. However, both of them fail to quantify how efficiently the energy is consumed, i.e., EE. Moreover, the optimality of EE and EC and that of EE and SE are not always achieved simultaneously and may even conflict with each other \cite{Funda_Tradeoff_YeLi}. As a consequence, the existing results from the EC minimization or the SE maximization usually can hardly provide insights into EE-SE tradeoff problems.


The general idea of modeling the EE-SE tradeoff is that the system maximizes the network EE \cite{EE_OFDMA_YeLi} or a weighted EE-SE tradeoff index \cite{ResourceEfficiency_OFDMA_TWC2014} under the constraints of users' QoS and resource allocation (e.g., power allocation and RRH operation). As a common feature, these works usually assume infinite backlog, i.e., there is always data for transmission in the buffer. Under this view, formulations are presented and algorithms are developed only based on the observation time, where the network EE is defined as the ratio of the instantaneous achievable sum rate $R_{\text{tot}}$ to the corresponding total power consumption $P_{\text{tot}}$ (cf. Eq.~(5) or (6a) in \cite{EE_OFDMA_YeLi}). Note that $P_{\text{tot}}$ is usually modeled to include both transmit and circuit energy consumption, which is affected by the power amplifier inefficiency, transmit power, and circuit power. In the article, we call these formulations short-term (i.e., snapshot-based) models, since only short-term system performance is considered. Accordingly, we denote the network EE of this kind of definition by $\text{EE}_{\text{short-term}}$ for simplicity.


Although there have been a large number of works to address the EE-SE tradeoff based on the short-term models, lots of problems remain open in complex H-CRANs. First, jointly considering multi-dimensional resource optimization and multi-available signal processing techniques, it is challenging to formulate EE-SE tradeoff problems with network conditions and users' requirements both taken into account in H-CRANs. Furthermore, due to the nonconvexity of $\text{EE}_{\text{short-term}}$ (cf. Eq.~(5) or (6a) in \cite{EE_OFDMA_YeLi} or Eq. (26) in \cite{ResourceEfficiency_OFDMA_TWC2014}), EE-SE tradeoff problems are usually difficult to solve even if we only optimize power allocation in spectrum-sharing H-CRANs. As a result, these problems become much more complicated once we extend from one-dimensional to multi-dimensional resource optimization. Thus, how to develop joint resource allocation algorithms that reach the theoretical limits of the network EE and thus serve as benchmarks to evaluate performance of other heuristic algorithms is another challenge. Moreover, it is also necessary to develop cost-efficient and easy-to-implement algorithms with acceptable performance levels to solve these problems for practical applications.

\subsection{EE-Fairness Tradeoff}
The widely studied EE-optimal problems (NEPs) in H-CRANs emphasize the network EE maximization without considering EE fairness, i.e., ignoring the EE of individual links. By purely benefiting the links in good network conditions (e.g., excellent wireless channel, little interference, low traffic loads, or all), the NEPs improve the network EE at the cost of the EE of the links in poor conditions. As a result, the NEPs would inevitably lead to severe unfairness among links in terms of EE. However, as traditional concerns on  individual links' SE or EC, it is also important to guarantee the EE of each link from users' perception. It is therefore of interesting to investigate the EE-fairness tradeoff in H-CRANs, but to the best of our knowledge, studies on this issue have so far been very scarce.

\begin{figure}[t]
\centering \leavevmode \epsfxsize=3.5in  \epsfbox{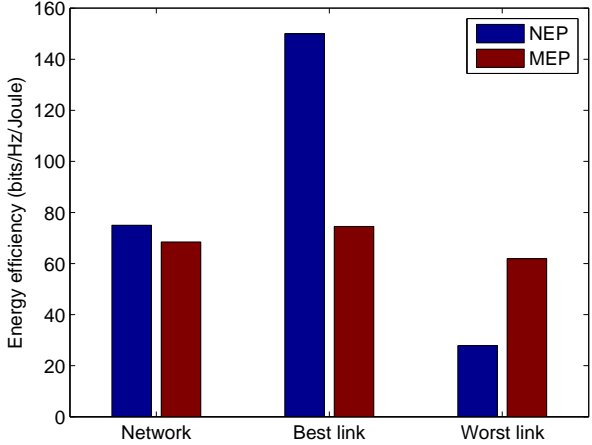}
\centering \caption{Illustration of the EE-fairness tradeoff. In this example, we consider an uplink OFDMA-based cellular network and formulate an optimization problem that maximizes the EE of its worst-case link subject to subcarrier assignment and power allocation constraints. In the figure, the number of users $K = 16$, number of subcarriers $N=128$, power amplifier inefficiency factor $\xi_k = 18$, terminal's circuit power $P_k^C = 0.4$ W, user's rate requirement $R_k^{\text{req}} = 15$ bits/s/Hz, and maximum transmit power $P_{k}^{\max} = 0.2$ W for all $k$. Note that the EE of the best/worst link is obtained by saving the EE of the link who has the highest/lowest EE in each sample and then taking an average on 5000 of them.} \label{Fig:EE_Network_Best_Worst_16_128}
\end{figure}

To intuitively show the EE-fairness tradeoff, we take the max-min EE fairness in an uplink OFDMA-based cellular network (it can be seen as a special case of single-cell H-CRANs) as an example. Specifically, we maximize the EE of the worst-case link subject to subcarrier assignment and power allocation constraints to ensure the max-min EE fairness among links, which is referred to as the max-min EE-optimal problem (MEP). In Fig.~\ref{Fig:EE_Network_Best_Worst_16_128}, we compare the statistical performance between the NEP and the MEP from three aspects: the EE of the network, the best link, and the worst link. Observe that, the EE of the best and worst links in the NEP differs significantly, while the EE whether of the network, the best link, or the worst link in the MEP is well-balanced. This is because the NEP maximizes the network EE at the cost of the EE fairness among links, but reversely, the MEP sacrifices the network EE to guarantee the max-min EE fairness.

Fig.~\ref{Fig:EE_Network_Best_Worst_16_128} exhibits the phenomenon of the EE-fairness tradeoff, but we are still at a very primary stage of revealing and tuning this tradeoff, limited by the following two challenges.
\begin{itemize}
  \item Unified frameworks to  quantify and formulate the EE-fairness tradeoff are currently not available.
  \item General techniques or analytical methods to tackle the EE-fairness tradeoff problems are still open.
\end{itemize}
\noindent It should be pointed out that the utility theory, originally used to investigate the rate-fairness tradeoff \cite{Fairness_Efficiency_Tradeoff_TON2013}, is a possible method to demystify the quantitative EE-fairness tradeoff.

\subsection{EE-Delay Tradeoff}
As far as we know, the concept of the EE-delay tradeoff was first proposed by H. V. Poor \emph{et al.} in 2009 \cite{EE_Delay_Tradeoffs_Game_IT2009}, where the authors showed that the delay constraints would lead to a loss in EE at equilibrium by a game-theoretical approach. However, to date, how to quantify and control the EE-delay tradeoff is still unresolved.

In our view, one possible reason that prevents the existing works including \cite{EE_Delay_Tradeoffs_Game_IT2009} from obtaining a quantitative tradeoff is the choice of adopting short-term models with the full buffer assumption, where $\text{EE}_{\text{short-term}}$ is used to characterize the network EE. However, different from the full buffer assumption, practical H-CRANs operate in the presence of time-varying wireless channels and stochastic traffic arrivals, both of which significantly affect the EE and delay and thus the EE-delay tradeoff. Hence, short-term formulations in general cannot reflect the delay due to their independence of time and without considering traffic arrivals. As a result, it is unlikely for such models to show the explicit EE-delay relationships.

We further illustrate the principles behind the EE-delay tradeoff with two extreme cases. Regarding stochastic traffic arrivals, in the case of aggressive emphasis on the EE, transmission decisions should be triggered only when network conditions are good enough, by which the delay performance degrades inevitably. Alternatively, to ensure a small delay, the network has to transmit data at the cost of energy expenditure even when network conditions are very poor, which undoubtedly decreases the EE. Thus, to model the EE-delay tradeoff, the following two issues need to be considered.
\begin{itemize}
  \item How to decide whether to transmit data or defer a transmission in each slot in terms of the EE and delay and how to optimize resource allocation such as power allocation, subcarrier assignment, and RRH operation if transmission is chosen?
  \item How to ensure that deferring transmissions to anticipate more advantageous network conditions becoming available in the future would not result in an uncontrollable delay because of time-variant, stochastic, and unpredicted network conditions?
\end{itemize}
In what follows, we present a possible method to model and reveal the quantitative EE-delay tradeoff.

To formulate EE and delay in a framework, we first need to shift from previously short-term to long-term models. In long-term formulations, random traffic arrivals can be enfolded to obtain a dynamic arrival-departure queue for each user, given as ${Q_i}\left( {t + 1} \right) = \max [{Q_i}\left( t \right) - {R_i}\left( t \right),0] + {A_i}\left( t \right), \forall i$ \cite{EE_Delay_D2D_JSAC2013}.
Here, ${A_i}\left( t \right)$ and ${Q_i}\left( t \right)$ denote the amount of newly arrived data and queue length of user $i$ at slot $t$, respectively. Note that the average delay can be characterized by queue length, as it is proportional to the queue length for a given traffic arrival rate from the Little's Theorem.

Furthermore, it is also necessary to inject the concept of time into the EE definition $\text{EE}_{\text{short-term}}$ in order to bridge the EE and delay. One possible way to achieve this is to define the EE from a long-term average perspective, given by the ratio of the long-term aggregate data delivered to the corresponding long-term total power consumption (cf. Eq.~(10) in \cite{EE_Delay_D2D_JSAC2013}). For simplicity, we denote this kind of the network EE definition by $\text{EE}_{\text{long-term}}$. From \cite{EE_OFDMA_YeLi} and \cite{EE_Delay_D2D_JSAC2013}, we know that, $\text{EE}_{\text{long-term}}$ can also be seen as an extension of $\text{EE}_{\text{short-term}}$, because it degenerates to $\text{EE}_{\text{short-term}}$ if there are no time averages and expectations in $\text{EE}_{\text{long-term}}$. Then, by integrating the queue length control (i.e., delay control) and EE maximization into a framework, we can depict the EE and average delay simultaneously.

\begin{figure}[t]
\centering \leavevmode \epsfxsize=3.5in  \epsfbox{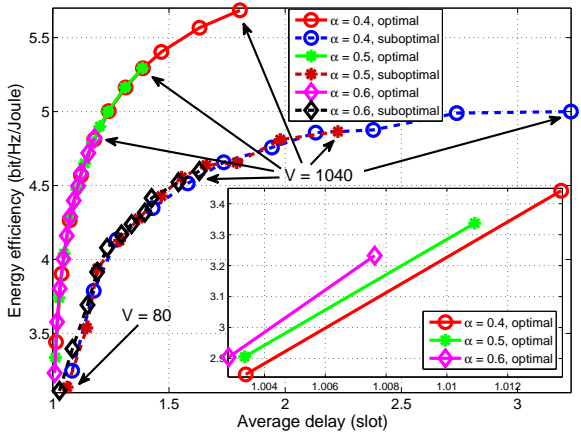}
\centering \caption{Illustration of the EE-delay tradeoff. In this example, we consider a downlink single-MBS H-CRAN and maximize its network EE $\text{EE}_{\text{long-term}}$ subject to a queue length control constraint by jointly optimizing RRH operation and power allocation. In the figure, the traffic arrival rate $\boldsymbol \lambda = 2.5$ bits/slot/Hz, RRH's circuit power $P_n^c = 0.4$ W, number of RRHs $N = 8$, and number of users $M = 12$. In particular, $V \ge 0$ and $\alpha \in [0,1]$ are two control parameters introduced to adjust the EE-delay tradeoff.}\label{Fig:EE_VS_Delay_ForAlpha}
\end{figure}

We utilize the above ideas to display the EE-delay tradeoff in H-CRANs by formulating a stochastic optimization problem that maximizes the network EE $\text{EE}_{\text{long-term}}$ subject to a queue length control constraint through joint optimization of RRH operation and power allocation. Two algorithms, referred to as the optimal and suboptimal, are developed to solve this problem. Fig.~\ref{Fig:EE_VS_Delay_ForAlpha} intuitively shows the EE-delay tradeoff, where $V \ge 0$ and $\alpha \in [0,1]$ are two control parameters introduced in the model to adjust the EE-delay tradeoff. Specifically, from Fig.~\ref{Fig:EE_VS_Delay_ForAlpha}, for the same $V$, the smaller $\alpha$ is, the better the EE, and the larger the average delay. In addition, for the same $\alpha$, the bigger $V$ is, the better the EE, and the larger the average delay. These observations together exhibit the EE-delay tradeoff, which can be explicitly balanced by $V$ and $\alpha$. Hence, the long-term model can be used to tune the EE-delay tradeoff via adjusting $V$ and $\alpha$. More clearly, $\alpha$ is used to confine the tradeoff range between the EE and average delay ( a small $\alpha$ gives a large range and vice versa) and $V$ to tune the tradeoff point between the EE and average delay (a small $V$ yields a small delay but low EE and vice versa).

Although \cite{EE_Delay_Tradeoffs_Game_IT2009} found the EE-delay tradeoff and \cite{EE_Delay_D2D_JSAC2013} obtained an EE-delay tradeoff of $[O\left(1/V\right),O\left(V\right)]$, the optimal EE-delay tradeoff, i.e., the optimal order for the average delay in $V$ when the EE increases to the optimal by the law of $O\left(1/V\right)$, is still unknown. Moreover, \cite{EE_Delay_Tradeoffs_Game_IT2009,EE_Delay_D2D_JSAC2013} focused on the average delay and thus the obtained results therein are valid only for non-real-time traffic such as web browsing and file transfers. However, there are some other real-time applications, e.g., voice and mobile video, in H-CRANs, which impose hard-deadline (or maximum delay) constraints. It is thus deserved to study how to provision deterministic delay guarantees and improve the EE in the meantime. Moreover, in more realistic H-CRANs with both non-real-time and real-time traffic, it is also well worth investigating how to flexibly balance the EE-delay performance for each kind of traffic from a perspective of systematic design and further devise control algorithms. Potential techniques that can be used to settle these unresolved issues are stochastic optimization, dynamic programming, Markov decision process, queue theory, and stochastic analysis.

\section{Conclusions} \label{Section:Conlusions}
Under the triple drives of capacity enhancement, EE improvement, and communication ubiquity, H-CRANs have emerged as a promising architecture for future wireless network design. In this article, we have first exploited the features of H-CRANs to propose three green techniques and then particularly focused on three fundamental tradeoffs, namely EE-SE, EE-fairness, and EE-delay tradeoffs. We have introduced the methods to model and analyze these tradeoffs, presented open issues and challenges, and also provided some potential solutions. However, we are still at a very primary stage in these studies, and thus further investigations on exploitation of the high-dimension, flexible, and scalable architecture of H-CRANs are eagerly deserved for a green future.

\section{Acknowledgement}
This work was supported in part by the National Science Foundation of China under Grant 61601192, 61601193, 61631015, 61471163, the U.S. NSF under Grant CNS-1320664, the Major State Basic Research Development Program of China (973 Program) under Grant 2013CB329006, the Major Program of National Natural Science Foundation of Hubei in China under Grant 2016CFA009, and the Fundamental Research Funds for the Central Universities under Grant 2016YXMS298.

\bibliographystyle{IEEEtran}
\bibliography{IEEEabrv,F://ReferenceOfPaper//MyRef}

\section*{Biographies}

{\footnotesize{\noindent Yuzhou~Li [M'14] (yuzhouli@hust.edu.cn) received the Ph.D. degree in communications and information systems from the School of Telecommunications Engineering, Xidian University, Xi'an, China, in December 2015. Since then, he has been with the School of Electronic Information and Communications, Huazhong University of Science and Technology, Wuhan, China, where he is currently an Assistant Professor. His research interests include 5G wireless networks, marine object detection and recognition, and undersea localization.}}

\vspace{1.3em}

{\footnotesize{\noindent Tao~Jiang [M'06-SM'10] (taojiang@hust.edu.cn) is currently a Distinguished Professor with the School of Electronics Information and Communications, Huazhong University of Science and Technology, Wuhan, P. R. China. He has authored or co-authored over 200 technical papers and 5 books in the areas of wireless communications and networks. He is the associate editor-in-chief of China Communications and on the Editorial Board of IEEE Transactions on Signal Processing, IEEE Transactions on Vehicular Technology, among others.}}

\vspace{1.3em}

{\footnotesize{\noindent Kai~Luo (kluo@hust.edu.cn) received his BEng degree from School of Electronics Information and Communications (EIC), Huazhong University of Science and Technology (HUST), China in 2006. Then, he received his Ph.D. degree in electrical engineering from Imperial College London in 2013. In 2013, he joined Institute of Electronics, Chinese Academy of Sciences. Since 2014, he is an Assistant Professor with School of EIC, HUST. His research interests are signal processing, MIMO communications and heterogeneous networks.}}

\vspace{1.3em}

{\footnotesize{\noindent Shiwen~Mao [S'99-M'04-SM'09] (shiwen.mao@gmail.com) received Ph.D. in electrical and computer engineering from Polytechnic University, Brooklyn, NY in 2004. He is the Samuel Ginn Distinguished Professor and Director of the Wireless Engineering Research and Education Center at Auburn University, Auburn, AL. His research interests include wireless networks and multimedia communications. He is a Distinguished Lecturer of the IEEE Vehicular Technology Society and on the Editorial Board of IEEE Transactions on Multimedia, IEEE Multimedia, among others.}}

\end{document}